\documentclass[prc,onecolumn,showpacs,superscriptaddress]{revtex4}
\usepackage{graphicx}
\begin{document}

\title{Calculations of the cross sections for synthesis of new $^{293-296}$118 isotopes
in $^{249-252}$Cf($^{48}$Ca,xn) reactions }

\author{ T. Cap}
\affiliation{Institute of Experimental Physics, Faculty of
Physics, University of Warsaw, Ho\.za 69, 00-681 Warsaw, Poland }

\author{K. Siwek-Wilczy\'nska}
\affiliation{Institute of Experimental Physics, Faculty of
Physics, University of Warsaw, Ho\.za 69, 00-681 Warsaw, Poland }

\author{M. Kowal}
\affiliation{National Centre for Nuclear Research, Ho\.za 69,
00-681 Warsaw, Poland   }

\author{J. Wilczy\'nski}
\affiliation{National Centre for Nuclear Research, 05-400
Otwock-\'Swierk, Poland   }

\date{\today}

\begin{abstract}

A project of using a target consisting of the mixture  of
$^{249-252}$Cf isotopes to be bombarded with the $^{48}$Ca beam,
aimed to synthesize new isotopes of the heaviest known element $Z$
= 118, is under way at the FLNR in Dubna. In the present work
excitation functions for all the reactions:
$^{249}$Cf($^{48}$Ca,xn)$^{297-x}$118,
$^{250}$Cf($^{48}$Ca,xn)$^{298-x}$118,
$^{251}$Cf($^{48}$Ca,xn)$^{299-x}$118 and
$^{252}$Cf($^{48}$Ca,xn)$^{300-x}$118 have been calculated in the
framework of the fusion-by-diffusion model, assuming fission
barriers, ground-state masses and shell effects of the superheavy
nuclei predicted by Kowal et al.  Energy dependence of the
effective cross sections for the synthesis of selected new
isotopes: $^{293}$118, $^{294}$118, $^{295}$118 and $^{296}$118 is
predicted for the particular isotopic composition of the Cf target
prepared for the Dubna experiment.
\end{abstract}


%\keywords{}

\maketitle

Following the great success of the synthesis of new superheavy
nuclei of $Z$ = 113-118 in hot fusion reactions, in which various
actinide targets ($^{242,244}$Pu, $^{243}$Am, $^{245,248}$Cm,
$^{249}$Bk and $^{249}$Cf) were bombarded with $^{48}$Ca
projectiles \cite{Oganessian-review}, a number of theoretical
attempts to reproduce the observed synthesis cross sections have
been made. Systematic model calculations were done for this class
of hot fusion reactions with the Langevin dynamics model
\cite{Zagrebaev-08}, fusion-by-diffusion model \cite{FBD-12} and
also with a phenomenological version of the di-nuclear system
(DNS) model \cite{Scheid-syst}. In Ref. \cite{Mandaglio},
calculations for the $^{48}$Ca induced reactions on $^{249-252}$Cf
targets have been done with the DNS model.

The aim of this Brief Report is to give predictions for the
experiment being under way at the Flerov Laboratory of Nuclear
Reactions in Dubna, in which a target consisting of a mixture of
$^{249-252}$Cf isotopes will be bombarded with the $^{48}$Ca beam
in order to synthesize new isotopes of the element $Z$ = 118. We
present the energy dependence of the evaporation-residue cross
sections for synthesis of $^{293}$118, $^{294}$118, $^{295}$118
and $^{296}$118 nuclides, predicted within the fusion-by-diffusion
(FBD) model \cite{FBD-Acta,FBD-05,FBD-11}. Below we give a short
description of this model.

As in other theoretical models used to describe synthesis of
superheavy nuclei, the partial evaporation residue cross section
$\sigma_{ER}(l)$ is factorized in the FBD model as the product of
the partial capture cross section $\sigma_{cap}(l)=\pi
\lambdabar^2(2l+1)T(l)$, the fusion probability $P_{fus}(l)$ and
the survival probability $P_{surv}(l)$:
\begin{equation}
\label{factorize}
 \sigma_{ER} = \pi \lambdabar^2 \sum_{l = 0}^{\infty}(2l+1)
 T(l)\, P_{fus}(l)\, P_{surv}(l).
\end{equation}
Here, $\lambdabar$ is the wave length, $\lambdabar^2=\hbar^2/2\mu
E_{c.m.}$, and $\mu$ is the reduced mass of the colliding system.
The capture transmission coefficients, $T(l)$, are calculated from
the sytematics of the fusion cross sections for lighter systems
\cite{FBD-11,KSW-04}.

The fusion probability $P_{fus}(l)$ is a key factor in all models
aimed to describe fusion of superheavy systems. It tells us what
is the probability that after reaching the capture configuration
(sticking), the colliding system will eventually overcome the
saddle point and fuse, avoiding reseparation. It is well known
that for very heavy and less asymmetric systems, $P_{fus}(l)$ is
much smaller than 1 and thus is responsible for the dramatically
small cross sections for the production of superheavy nuclei. The
fusion hindrance in these reactions is caused by the fact that for
the heaviest compound nuclei the saddle configuration is more
compact than the configuration of the two initial nuclei at
sticking. It is assumed in the FBD model that after the sticking,
a neck between the two nuclei grows rapidly at an approximately
fixed mass asymmetry and constant length of the system
\cite{FBD-Acta,FBD-05} bringing the system to the ``injection
point'' somewhere along the bottom of the asymmetric fission
valley. To overcome the saddle point and fuse, the system must
climb uphill from the injection point to the saddle in the process
of thermal fluctuations in shape degrees of freedom. (A similar
scenario of fusion of the heaviest nuclear systems has been
demonstrated analytically in a simple two-dimensional Langevin
dynamics model \cite{Boilley}.) The location of the injection
point, $s_{inj}$, is the only adjustable parameter of the FBD
model. It was shown in Ref.~\cite{FBD-Acta} that by solving the
Smoluchowski diffusion equation, the probability that the system
injected on the outside of the saddle point at an energy $H$ below
the saddle point will achieve fusion is
\begin{equation}
\label{hindrance} P_{fus} = \frac{1}{2}(1 - {\rm
erf}\sqrt{H/T}\,),
\end{equation}
where $T$ is the temperature of the fusing system. The energy
threshold $H$ opposing fusion in the diffusion process is thus the
difference between the energy of the saddle point and the energy
of the combined system at the injection point, calculated using
the algebraic approximate expressions given in Ref. \cite{FBD-11}.
The corresponding values of the rotational energy at the injection
point and at the saddle point are calculated assuming the
rigid-body moments of inertia at these configurations.

The last factor in Eq.~(\ref{factorize}), $P_{surv}(l)$, is the
probability for the compound nucleus to decay to the ground state
of the residual nucleus via evaporation of light particles
(neutrons) and thus avoid fission (survive). To calculate the
survival probability $P_{surv}$, the standard statistical model
was used by applying the Weisskopf formula for the neutron
emission width $\Gamma_n$ and the standard expression of the
transition-state theory for the fission width $\Gamma_f$. The
level density parameters $a_n$ and $a_f$ for neutron evaporation
and fission channels were calculated as proposed by Reisdorf
\cite{Reisdorf} with shell effects accounted for by the Ignatyuk
formula \cite{Ignatyuk}. All details can be found in Ref.
\cite{FBD-11}. In case of calculating multiple evaporation ($xn$)
channels a simplified algorithm avoiding the necessity of using
the Monte Carlo method was used \cite{Cap-xn}.

As it follows from the above description, the cross section
calculations require to know individual characteristics of the
synthesized compound nuclei and their decay products, first of
all, the fission barriers, ground-state masses and shell effects
as well as deformations of the compound nuclei in the ground state
and the saddle-point configuration. It was demonstrated in Ref.
\cite{FBD-12} that the fission barriers and other characteristics
of superheavy nuclei calculated according to the Warsaw
macroscopic-microscopic model \cite{Kowal-10,Kowal-arXiv} have
proved to well reproduce cross sections of hot (xn) fusion
reactions leading to the synthesis of $Z$ = 114--118 superheavy
nuclei. The recent Warsaw group calculations have been done in
multidimensional deformation space including nonaxial and
reflection-asymmetric shapes. As tables in Refs.
\cite{Kowal-10,Kowal-arXiv} are limited to even-even nuclei, the
fission barrier heights for the odd-$A$ nuclei have been
calculated separately \cite{Kowal-odd,Kowal-private} by adding the
energy of the odd particle occupying a single-particle state.

\begin{figure}[t]
\begin{tabular}{cc}
\resizebox{75mm}{!}{\includegraphics[angle=0,scale=0.75]{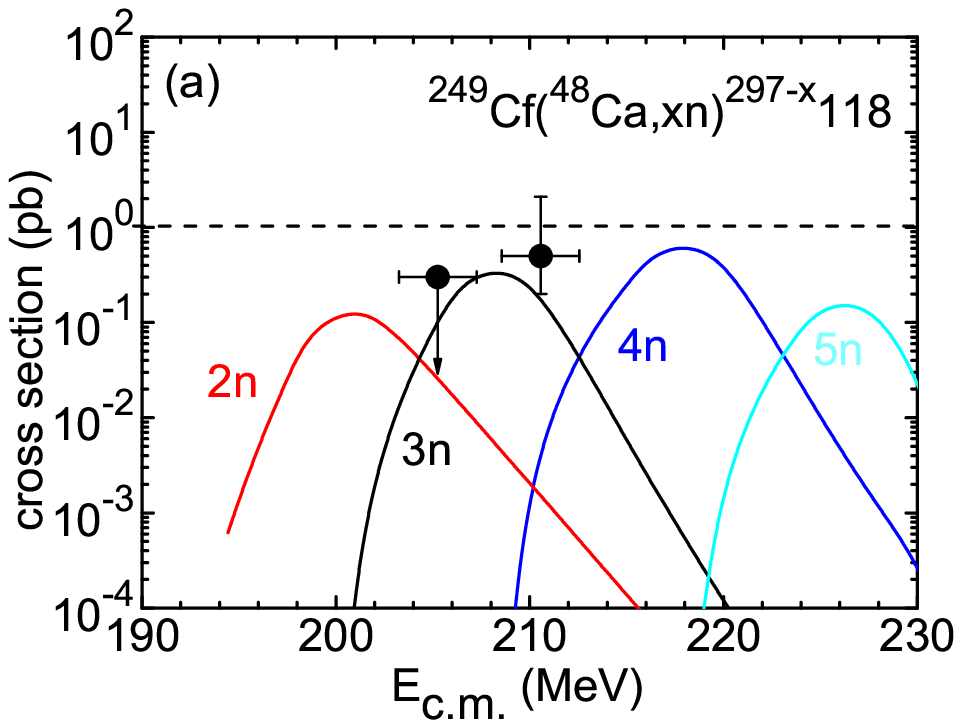}}
 &
\resizebox{75mm}{!}{\includegraphics[angle=0,scale=0.75]{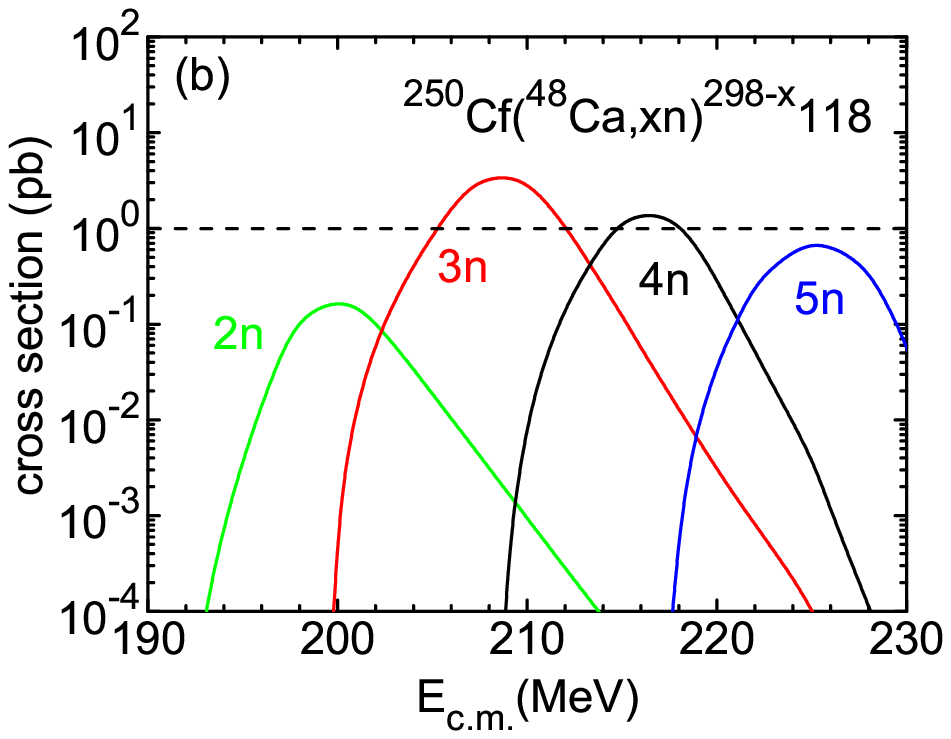}}\\

\resizebox{75mm}{!}{\includegraphics[angle=0,scale=0.75]{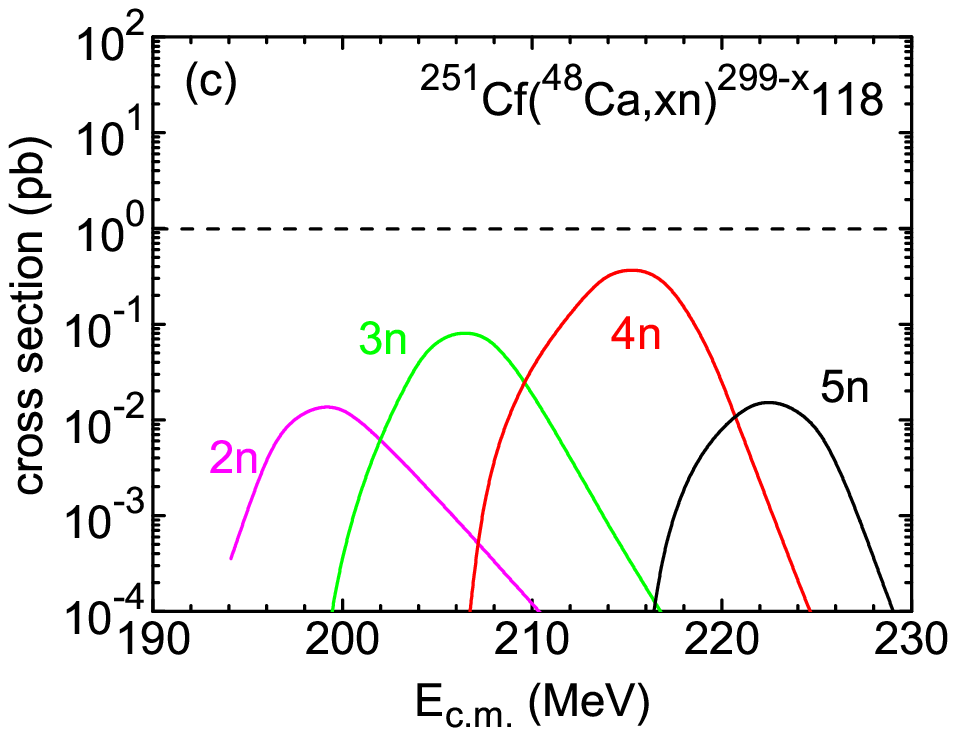}}
 &
\resizebox{75mm}{!}{\includegraphics[angle=0,scale=0.75]{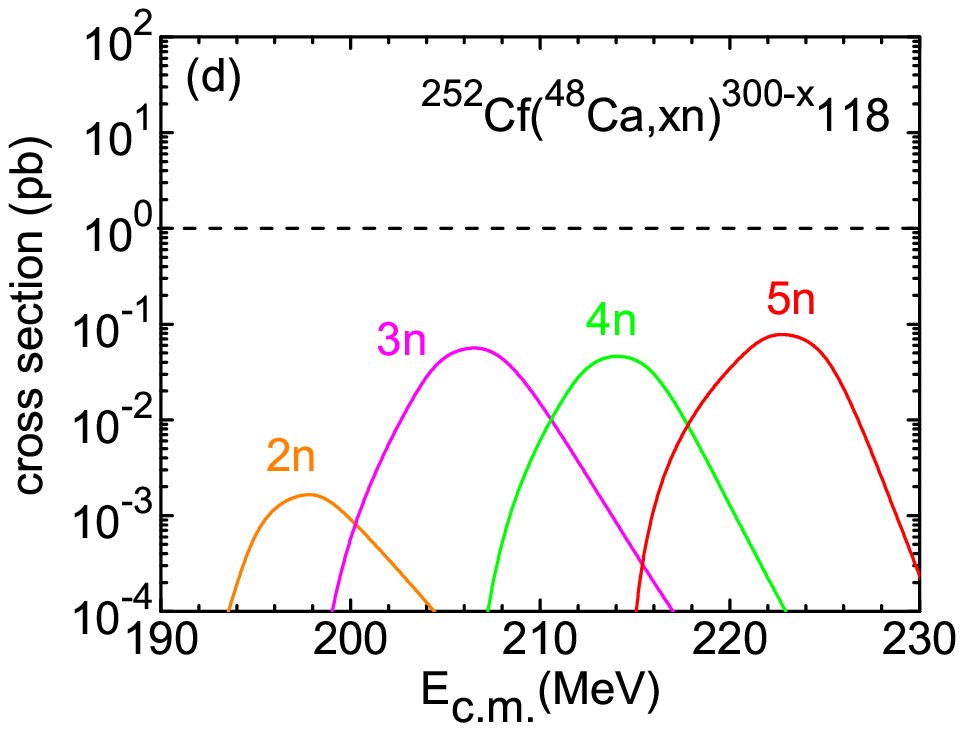}}\\

\end{tabular}
\caption{(Color online) Energy dependence of the
evaporation-residue cross sections in the
$^{249-252}$Cf($^{48}$Ca,xn) reactions on four separate Cf
isotopes, leading to the synthesis of various isotopes of the
element $Z$ = 118. The excitation functions have been calculated
within the fusion-by-diffusion model, assuming theoretical fission
barriers of Kowal et al.
\cite{Kowal-10,Kowal-arXiv,Kowal-odd,Kowal-private}. The
calculated excitation functions have been corrected to account for
the effective beam energy spread due to the beam energy loss in
the target of 4 MeV. Curves corresponding to the same final
isotope produced in different reactions are drawn the same color,
also in Fig. 2. }
\end{figure}

As mentioned above, the heaviest new element with atomic number
118 was synthesized in the $^{249}$Cf($^{48}$Ca,3n)$^{294}$118
reaction \cite{Oganessian-review,Z=118}. This particular
experiment was carried out by using a nearly mono-isotopic
$^{249}$Cf target of $>$98$\%$ purity \cite{Z=118}. Recently, a
project to produce a target consisting of the mixture of
$^{249-252}$Cf isotopes for experiments  with the $^{48}$Ca beam
was proposed \cite{Oganessian-private}. By using this target,
there will be a chance to synthesize more isotopes of element 118,
in addition to the $^{294}$118 nuclide that was produced in the
earlier experiment \cite{Z=118}. The synthesis efficiency in this
experiment is expected to be widened to a larger number of
isotopes of element 118 as a result of simultaneous production of
different isotopes in several xn channels on different Cf isotopes
present in the target mixture. The isotopic content of the
prepared target material is $^{249}$Cf (42.31\%), $^{250}$Cf
(21.76\%), $^{251}$Cf (35.64\%) and $^{252}$Cf (0.29\%)
\cite{Oganessian-private}.

In the present work we calculated excitation functions for the
reactions with all four Cf isotopes present in the target:
$^{249}$Cf($^{48}$Ca,xn)$^{297-x}$118,
$^{250}$Cf($^{48}$Ca,xn)$^{298-x}$118,
$^{251}$Cf($^{48}$Ca,xn)$^{299-x}$118 and
$^{252}$Cf($^{48}$Ca,xn)$^{300-x}$118. The calculations have been
done exactly according to the scheme \cite{FBD-12} used for
analysis of the whole set of hot fusion reactions leading to the
synthesis of elements $Z$ = 114-118, with fission barriers and
other theoretical characteristics of the superheavy compound
nuclei \cite{Kowal-10,Kowal-arXiv,Kowal-odd,Kowal-private}, and
systematics of the injection point distance determined in
\cite{FBD-12}.

The calculated excitation functions are displayed in Fig. 1.
Figure 1a shows the results for the $^{48}$Ca + $^{249}$Cf
reaction, in which the $^{294}$118 isotope of element 118 was
observed \cite{Z=118} in the 3n reaction channel. Along with the
calculated excitation functions, the experimental value of the
cross section evaluated at $E_{c.m.}$ $\approx$ 210 MeV and the
upper limit of the 3n cross section at $E_{c.m.}$ $\approx$ 205
MeV are shown. For easier comparisons, dashed lines in all figures
show the 1 pb level of the cross section that is a typical limit
of sensitivity in modern experiments aimed to synthesize
superheavy elements. It should be noted that in addition to the 3n
reaction, our calculations for the $^{48}$Ca + $^{249}$Cf reaction
predict a sizable cross section of the 4n reaction leading to the
synthesis of the $^{293}$118 nuclide. The maximum of the
theoretical 4n excitation function is located at $E_{c.m.}$
$\approx$ 217 MeV, that is well above the range of energies
covered in the experiment \cite{Z=118}.

As seen from Fig. 1b, the reaction on the $^{250}$Cf isotope gives
a chance to observe the new nuclide $^{295}$118 expected to be
produced with relatively large cross section of about 3 pb in the
3n reaction at $E_{c.m.}$ $\approx$ 208 MeV. Also the $^{294}$118
isotope is expected to be produced in the $^{48}$Ca + $^{250}$Cf
reaction  with a measurable cross section of about 1 pb in the 4n
channel. The maximum of the 4n excitation function is predicted at
$E_{c.m.}$ $\approx$ 216 MeV.

\begin{figure}[t]
\begin{tabular}{cc}
\resizebox{75mm}{!}{\includegraphics[angle=0,scale=0.75]{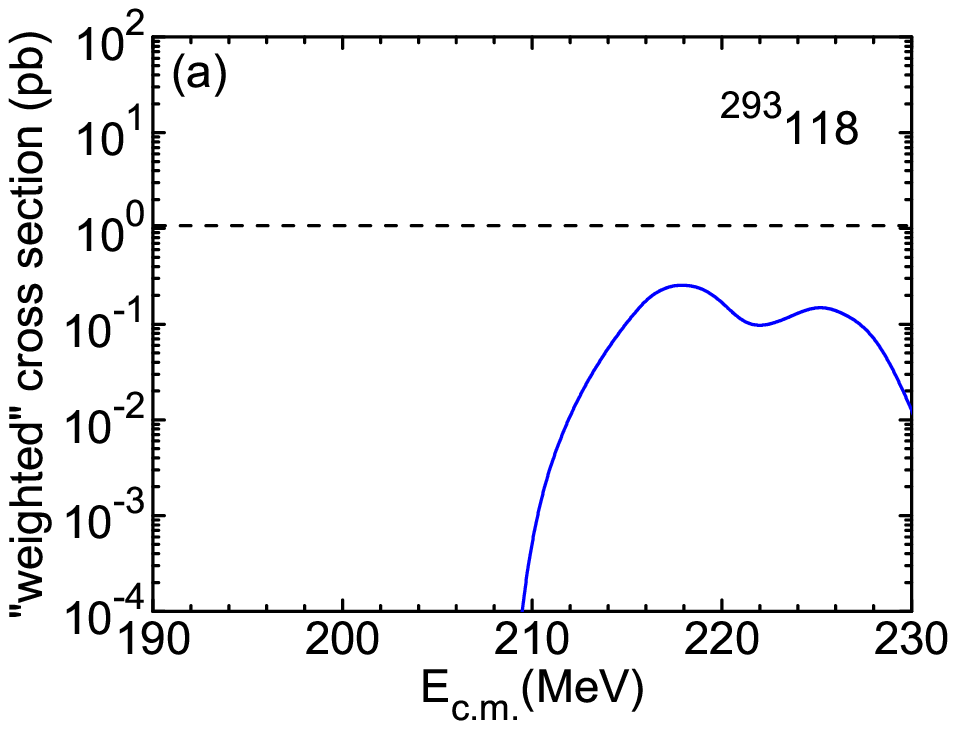}}
 &
\resizebox{75mm}{!}{\includegraphics[angle=0,scale=0.75]{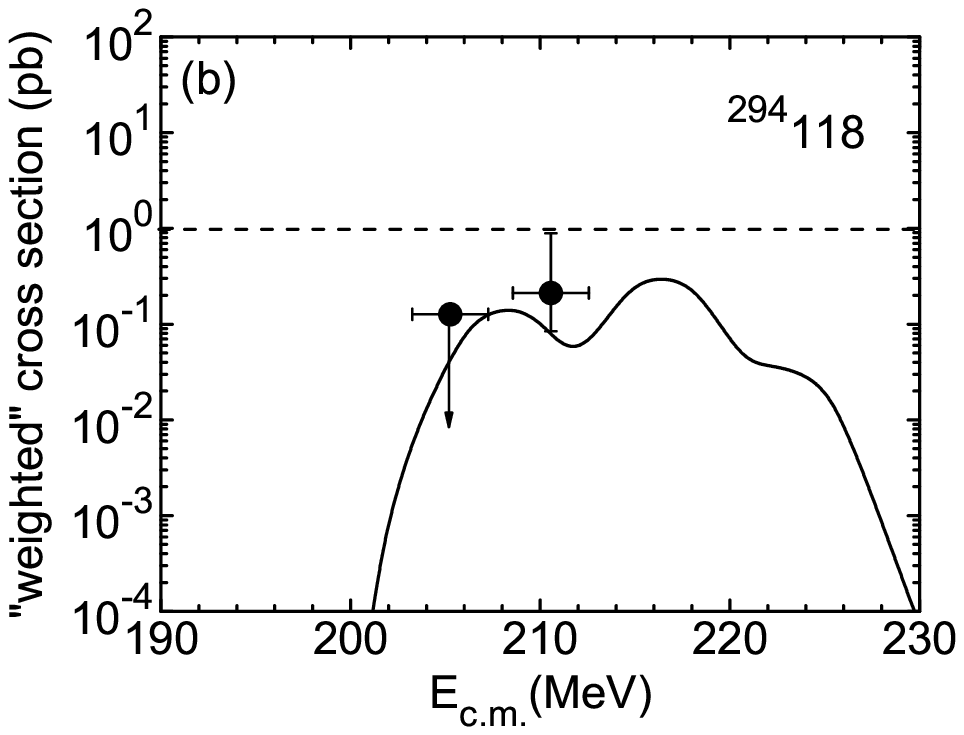}}\\

\resizebox{75mm}{!}{\includegraphics[angle=0,scale=0.75]{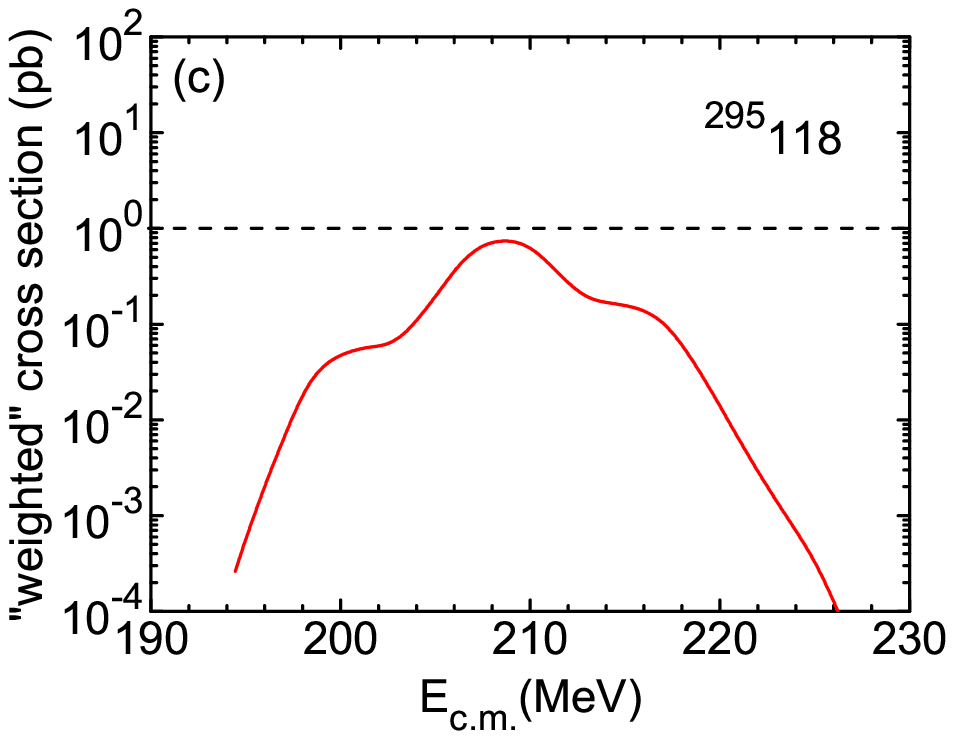}}
 &
\resizebox{75mm}{!}{\includegraphics[angle=0,scale=0.75]{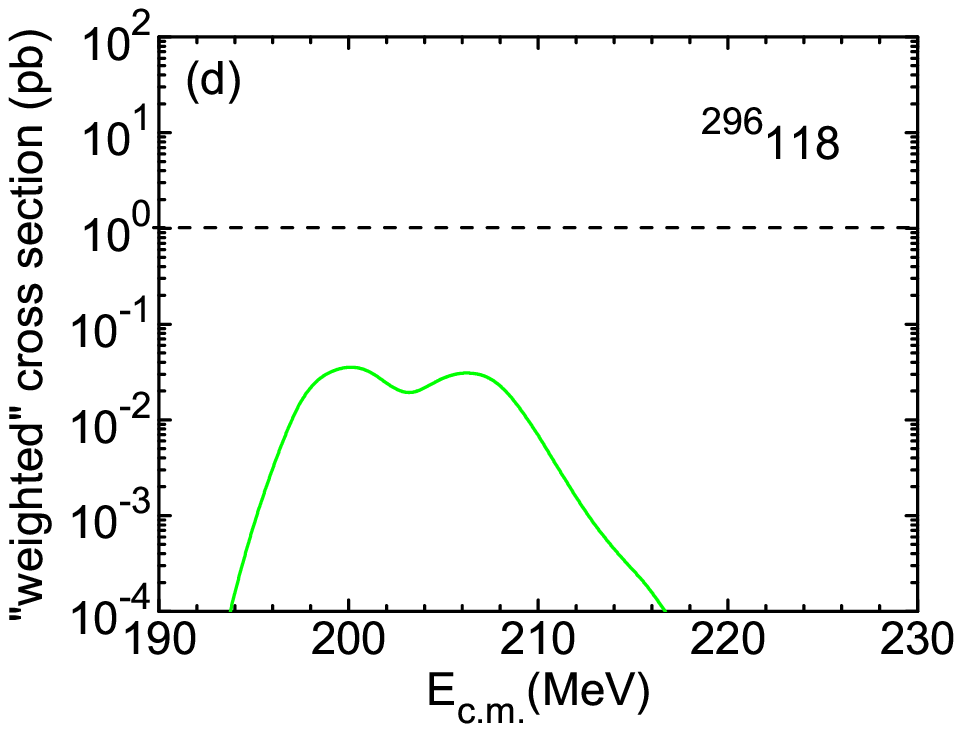}}\\

\end{tabular}
\caption{(Color online) Energy dependence of the {\em weighted}
cross sections for the synthesis of separate isotopes $^{293}$118,
$^{294}$118, $^{295}$118 and $^{296}$118 in various channels of
the $^{249-252}$Cf($^{48}$Ca,xn) reactions, predicted for the
isotopic composition of the Cf target as planned to be used in the
experiment at FLNR in Dubna: 42.31\% of $^{249}$Cf, 21.76\% of
$^{250}$Cf, 35.64\% of $^{251}$Cf and 0.29\% of $^{252}$Cf
\cite{Oganessian-private}. The data points in Fig. 2(b) refer to
the $^{249}$Cf($^{48}$Ca,3n) reaction [see Fig. 1(a)]. They are
rescaled accordingly with the definition of the ``{\em weighted}
cross section''.}
\end{figure}

Our predictions show, unfortunately, that the use of the most
neutron rich isotopes, $^{251}$Cf and $^{252}$Cf,  is not a
perspective way for synthesis of new isotopes of the element 118.
As seen from Figs. 1c and 1d, the expected cross sections become
extremely small, of the order of 0.1 pb, due to the decreasing
theoretical values of the fission barrier with the increasing
neutron number for the heaviest $Z$ = 118 isotopes.

Figure 2 displays the excitation functions for the synthesis of
separate isotopes $^{293}$118, $^{294}$118, $^{295}$118 and
$^{296}$118, calculated for the mixture $^{249-252}$Cf target,
taking the actual content of each isotope in the mixture. The
``{\em weighted} cross section'' displayed in the diagrams is the
sum of the synthesis cross section for a given final isotope of
element 118 produced on all four Cf isotopes in the corresponding
xn reaction channels, reduced by the factor of relative content of
the target isotope in the mixture. For example, the {\em weighted}
cross section for synthesis of the $^{294}$118 nucleus is:
$\sigma$($^{294}$118) =
$f_1\sigma$[$^{249}$Cf($^{48}$Ca,3n)$^{294}$118] +
$f_2\sigma$[$^{250}$Cf($^{48}$Ca,4n)$^{294}$118] +
$f_3\sigma$[$^{251}$Cf($^{48}$Ca,5n)$^{294}$118] +
$f_4\sigma$[$^{252}$Cf($^{48}$Ca,6n)$^{294}$118], where $f_1$ =
0.4231, $f_2$ = 0.2176, $f_3$ = 0.3564 and $f_4$ = 0.0029
\cite{Oganessian-private}. In order to easier identify separate
components of the yield of a given isotope presented in Fig. 2,
the respective xn curves in Fig. 1 are drawn the same color.

As one can see from Fig. 2, the largest {\em weighted} cross
section, close to 1 pb, is expected for the production of the
$^{295}$118 nucleus at $E_{c.m.}$ $\approx$ 208 MeV (see Fig. 2c).
The half life of $^{295}$118 is expected \cite{Kowal-private} to
be longer than that of the already detected isotope $^{294}$118,
enabling experimental identification of the new isotope of element
118. Its decay chain should follow the decay of the already
discovered $^{291}$Lv nucleus.

A possibly measurable cross section of few hundreds fb is expected
for the unknown yet isotope $^{293}$118 (see Fig. 2a). In this
case the maximum cross section is predicted around $E_{c.m.}$ =
218 MeV. Therefore an attempt to synthesize the $^{293}$118
isotope would probably require to carry out a separate run at an
energy $E_{c.m.}$ = 216--218 MeV, in addition to the main
experiment at $E_{c.m.}$ = 208--210 MeV focused on the synthesis
of the $^{294}$118 and $^{295}$118 nuclei. Possible discovery of
the $^{293}$118 isotope would be extremely valuable because its
$\alpha$ decay product, $^{289}$Lv, is also unknown and further
decay products, $^{285}$Fl and $^{281}$Cn, would be important
cross checks of the present boundary of nuclides synthesized in
hot fusion reactions. Theoretical calculations
\cite{Kowal-private} predict that the life-times of the
$^{293}$118 nucleus and of its $\alpha$ decay products are long
enough to be detected in the experiment.

In summary, we calculated excitation functions of the synthesis of
isotopes of the element 118 in  $^{249-252}$Cf($^{48}$Ca,xn)
reactions. The calculations have been done within the
fusion-by-diffusion model, with fission barriers and other
theoretical characteristics of the superheavy compound nuclei
calculated with the Warsaw macroscopic-microscopic model.
Anticipating implementation of the Dubna experiment with the
target consisting of the mixture of the $^{249-252}$Cf isotopes
\cite{Oganessian-private}, calculations of the excitation
functions for the synthesis of separate isotopes $^{293}$118,
$^{294}$118, $^{295}$118 and $^{296}$118 have been done for the
particular isotopic composition of the Cf target prepared for this
experiment. The calculations predict observation of the already
known nuclide $^{294}$118 with the {\em weighted} synthesis cross
section of about 0.2--0.3 pb at $E_{c.m.}$ = 208--216 MeV, and
also the new nuclide $^{295}$118 with a larger {\em weighted}
cross section of about 1 pb at the bombarding energy $E_{c.m.}
\approx 208$ MeV. There is also a chance to synthesize another new
nuclide $^{293}$118 (and its unknown decay product $^{289}$Lv)
with the {\em weighted} cross section of about 0.2 pb at
$E_{c.m.}$ = 216--218 MeV.


\begin{thebibliography}{99}

\bibitem{Oganessian-review}
Yu. Oganessian, J. Phys. G: Nucl. Part. Phys. {\bf 34}, R165
(2007).

\bibitem{Zagrebaev-08}
V. Zagrebaev and W. Greiner, Phys. Rev. C {\bf 78}, 034610 (2008).

\bibitem{FBD-12}
K. Siwek-Wilczy\'nska, T. Cap, M. Kowal, A. Sobiczewski, and J.
Wilczy\'nski, Phys. Rev. C {\bf 86}, 014611 (2012).

\bibitem{Scheid-syst}
Ning Wang, Junlong Tian, and Werner Scheid, Phys. Rev. C {\bf 84,}
061601(R) (2011).

\bibitem{Mandaglio}
G. Mandaglio, G. Giardina, A. K. Nasirov, and A. Sobiczewski,
Phys. Rev. C {\bf 86,} 064607 (2012).

\bibitem{FBD-Acta}
W.~J.~\'Swi\c{a}tecki, K.~Siwek-Wilczy\'nska, and J.~Wilczy\'nski,
{\it Acta Phys. Pol.} {\bf B34}, 2049 (2003).

\bibitem{FBD-05}
W. J. \'Swi\c{a}tecki, K. Siwek-Wilczy\'nska, and J. Wilczy\'nski,
Phys. Rev. C {\bf 71,} 014602 (2005).

\bibitem{FBD-11}
T. Cap, K. Siwek-Wilczy\'nska and J. Wilczy\'nski, Phys. Rev. C
{\bf 83,} 054602 (2011).

\bibitem{KSW-04}
K. Siwek-Wilczy\'nska and J. Wilczy\'nski,  Phys. Rev. {\bf C69,}
024611 (2004).

\bibitem{Boilley}
David Boilley, Hongliang L\"u, Caiwan Shen, Yasuhisa Abe, and
Bertrand G. Giraud, Phys. Rev. C {\bf 84}, 054608 (2011).

\bibitem{Reisdorf}
W.~Reisdorf, {\it Z. Phys.} {\bf A300}, 227 (1981).

\bibitem{Ignatyuk}
A.~V.~Ignatyuk, G.~N.~Smirenkin, and A.~S.~Tishin, {\it Yad. Fiz.}
{\bf 21}, 485 (1975) [ {\it Sov. J. Nucl. Phys.} {\bf 21}, 255
(1975)].

\bibitem{Cap-xn}
T. Cap, K. Siwek-Wilczy\'nska, I. Skwira-Chalot, and J.
Wilczy\'nski, Acta Phys. Pol. {\bf B43,} 297 (2012).

\bibitem{Kowal-10}
M. Kowal, P. Jachimowicz, and A. Sobiczewski, Phys. Rev. C {\bf
82,} 014303 (2010).

\bibitem{Kowal-arXiv}
M. Kowal, P. Jachimowicz, and J. Skalski, arXiv:1203.5013

\bibitem{Kowal-odd}
P. Jachimowicz, M. Kowal, and J. Skalski, to be published.

\bibitem{Kowal-private}
M. Kowal, private communication.

\bibitem{Z=118}
Yu. Ts. Oganessian {\it et al.,} Phys. Rev. C {\bf 74}, 044602
(2006).

\bibitem{Oganessian-private}
Yu. Ts. Oganessian, private communication.

\end{thebibliography}
\end{document}